\documentclass[a4paper,11pt]{article}
\usepackage{pos}

\title{Axion-like Particle Conversion in the Solar Magnetic Field}

\author{Elisa Todarello}

\affiliation{School of Physics and Astronomy, University of Nottingham,
University Park, NG7 2RD, Nottingham, United Kingdom}

\emailAdd{elisa.todarello@nottingham.ac.uk}

\abstract{Axion-like particles (ALPs), hypothetical extensions of the Standard Model, can convert into photons in an external magnetic field. Two recent studies~\cite{Todarello:2023ptf, Ruz:2024gkl} explored the phenomenology of ALP-photon conversion in the magnetic field of the solar atmosphere. Dark matter ALPs convert into radio photons and may be detected with next-generation radio interferometers, while ALPs produced in the solar core convert into X-rays. Thanks to solar observation acquired with the NuSTAR X-ray telescope, Ref.~\cite{Ruz:2024gkl}  establishes stringent robust bounds on the ALP-photon coupling over a large portion of parameter space.}

\FullConference{2nd Training School and General Meeting of the COST Action COSMIC WISPers  (CA21106) (COSMICWISPers2024)\\
10-14 June 2024 and 3-6 September 2024\\
Ljubljana (Slovenia) and Istanbul (Turkey)\\}

\begin{document}
\maketitle

%%%%%%%%%%%%%%%%%%%%%%%%%%%%%%%%%%%%
\section{Introduction}

Axion-like particles (ALPs) are hypothetical bosons similar to the QCD axion~\cite{Peccei:1977hh,Peccei:1977ur, Weinberg:1977ma, Wilczek:1977pj} and can constitute the Dark Matter (DM) of our Universe~\cite{Arias:2012az}. ALPs and QCD axions share the characteristic interaction with photons that allows for conversions between axions/ALPs and photons in the presence of strong electromagnetic fields.

The probability of conversion of an ALP into a photon can be enhanced in the presence of a plasma. The plasma frequency acts as an effective photon mass, allowing for resonant conversion at locations where the dispersion relations of ALPs and photons match. In weakly magnetized plasmas, this condition is equivalent to the plasma frequency being equal to the ALP mass.

Thanks to its magnetic fields, plasma, and proximity, the Sun is an ideal environment to study such conversions. Highly relativistic ALPs are produced in the solar core and can convert into X-ray photons in the solar atmosphere. Similarly, DM ALPs can convert into radio waves in the magnetic field of the solar corona.
Here, we report on two recent works examining the ALP-photon conversion process in the Sun's atmosphere and its implications for relativistic solar ALP~\cite{Ruz:2024gkl} and non-relativistic ALP dark matter~\cite{Todarello:2023ptf}. 

%%%%%%%%%%%%%%%%%%%%%%%%%%%%%%%%%%%%
\section{ALP-Photon Conversion}
ALPs interact with electromagnetic fields via the effective Lagrangian term
\begin{equation}
    \mathcal{L}=g\,a\, \vec{E}\cdot\vec{B}\enspace,\label{eq:lagr}
\end{equation}
where $g$ is the ALP-photon coupling constant, $a$ is the ALP field, and $\vec{E}$ and $\vec{B}$ are the electric and magnetic field, respectively. This interaction allows ALP decay into photons and ALP-photon conversion, also called the Primakoff effect.
In a static background, ALP and photon will have the same energy $\omega$.
Let us further assume that the ALP and photon propagate in the same direction. Then, the probability of ALP to photon conversion is (see for example~\cite{Leroy:2019ghm, Sikivie:2020zpn})
\begin{equation}
    P_{a\rightarrow\gamma}(h) = \frac{1}{4}g^2 \frac{1}{v(h)} \ \Big|\int^h dh' \frac{1}{\sqrt{n(h')}}\,B_\perp(h')\ e^{i\int^{h’}dh''q(h'')}\Big|^2 \enspace,\label{eq:prob}
\end{equation}
where $v$ is the velocity of the ALP, $h$ is the distance traveled, $n$ is the index of refraction, and $B_\perp$ is the component of the magnetic field perpendicular to the direction of propagation. The factor appearing in the phase represents the momentum mismatch between ALPs with mass $m$ and photons
\begin{equation}
    q = k - k_a = n\omega - \sqrt{\omega^2 - m^2}\enspace.\label{eq:q}
\end{equation}

Although ALP and photon have the same energy, their momenta $k$ and $k_a$ are, in general, different. In a weakly magnetized plasma, like that of the solar corona, the index of refraction takes the simple form $n = \sqrt{\omega^2 - \omega_p^2}/\omega$, where $\omega_p$ is the plasma frequency. We see from~\eqref{eq:q} that the complex phase vanishes at locations where $\omega_p = m$, where resonant conversion can happen.

%%%%%%%%%%%%%%%%%%%%%%%%%%%%%%%%%%%%
\section{The Sun}
The solar atmosphere is customarily divided into layers. The lower layer is the photosphere, where the matter density drops to values such that the Sun becomes transparent to visible light. More precisely, the visible surface of the Sun is defined by the condition that the optical depth for green light be $\tau=2/3$. 
The photosphere is characterized by granules, the top edges of convection cells where hot material rises and cool material sinks. Magnetic field lines emerge from the solar surface along the lanes between granules and supergranules, which also host convection but on a larger scale, as illustrated, for example, in~\cite{Wedemeyer-Bohm:2008tpk}. 

The next layer is the chromosphere, extending approximately from 500 to 2,000 km above the surface of the Sun. Unlike in the photosphere, in this region, the temperature increases with altitude. Above the chromosphere, there is the transition region, a thin layer in which the temperature rises abruptly from $\sim 10^4$ to $10^6$~K, with one of the steepest gradients in the solar atmosphere. 

The outer layer of the solar atmosphere is the corona, characterized by temperatures of order 1 million kelvins. The origin of such high temperatures is still not fully understood~\cite{2024arXiv240913318A}.
The corona eventually becomes the solar wind, extending to distances well beyond Earth's orbit.

The Sun goes through cycles, each one lasting approximately 11 years. The active phase of the cycle is characterized by the inversion of the magnetic poles and by localized phenomena such as sunspots, solar flares, and coronal mass ejections.
Sunspots are dark areas on the Sun's surface caused by intense magnetic activity. They are cooler than the surrounding regions and can host magnetic fields with strength up to thousands of Gauss. The largest sunspots have sizes up to $\sim 10^4$~km.

During the quiet phase, on the other hand, the magnetic field in the solar corona has a typical strength of order 1~G. Both the quiet Sun's and sunspot's magnetic fields are interesting environments to study ALP-photon conversions.

In addition to being an ``ALP-converter", the Sun is also an ALP source. 
ALPs are produced in the solar core thanks to their interactions with photons and fermions. These ALPs emerge from the solar surface along approximately radial trajectories and with energies of order the temperature of the core. The emission is concentrated in the central part of the solar disc. In Section~\ref{sec:nustar}, we consider ALPs produced by the Primakoff conversion of thermal photons in the electric field of electrons and ions in the core. Such ALPs have a characteristic spectrum, peaked at an energy of about 3~keV.

%%%%%%%%%%%%%%%%%%%%%%%%%%%%%%%%%%%%
\section{Conversion of Dark Matter ALPs}
\begin{figure}[th]
    \centering
    \includegraphics[width=0.6\textwidth]{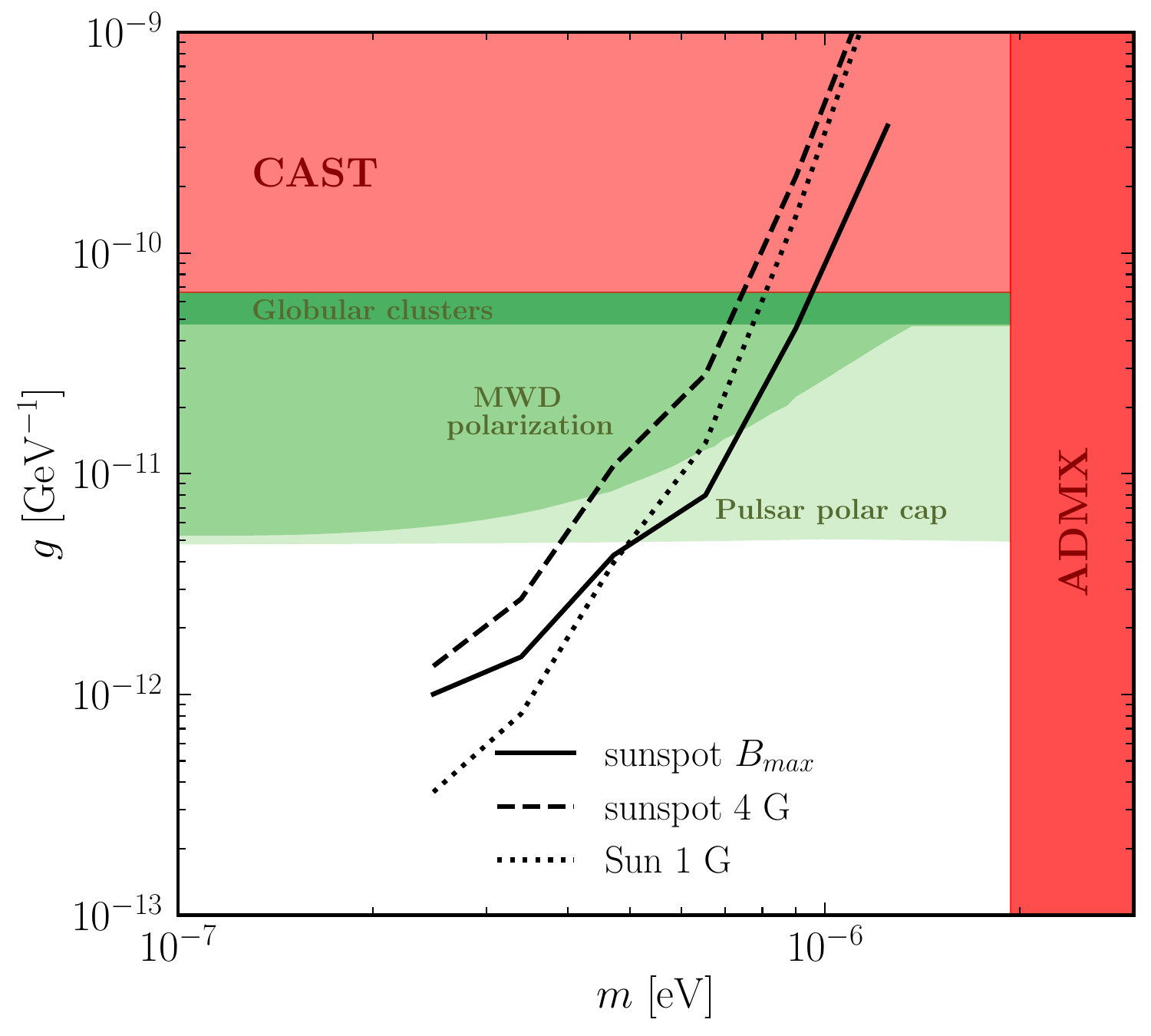}
    \caption{Projected sensitivity on the ALP-photon coupling $g$ versus ALP mass $m$, assuming 100 hours of observations with SKA-1 Low.
The solid line represents the sensitivity for observing a sunspot with the maximum magnetic field allowed by the gyro-resonance absorption condition $\omega_p > n\Omega_B$ at the conversion location, with $k=4$ (see~\cite{Todarello:2023ptf} for more details). The radius of the conversion surface above the sunspot is assumed to be $\ell_s = 4 \times 10^4$ km. The dashed line corresponds to observations of a sunspot of the same size but with a magnetic field of 4~G, while the dotted line represents the sensitivity for the whole quiet Sun.  
Laboratory bounds from ADMX~\cite{Asztalos_2010} and CAST~\cite{CAST2017} are shown in red, and astrophysical bounds are indicated~\cite{Dolan_2022,Dessert_2022,noordhuis2023novel} in green. Figure adapted from~\cite{Todarello:2023ptf}.
   }
    \label{fig:res}
\end{figure}

In the case of DM ALPs, the signal is a narrow spectral line centered around the average ALP energy $\omega\approx m$. The width of the spectral line is given by the DM energy dispersion $\delta\omega\sim 10^{-6}\,m$.
The probability of conversion is dominated by the contribution at the resonance location $h_{res}$ and can be expressed as
\begin{equation}
P_{a\rightarrow\gamma} \simeq \frac{\pi}{2}\frac{g^2\,B_{\bot }^2}{v \, |\omega_p^{\prime}|}\Big|_{h=h_{res}}\enspace.
\end{equation}
Here, $\omega_p^{\prime}$ is the derivative of the plasma frequency along the direction of propagation. Given the typical electron density of the solar corona $n_e$, and $\omega_p= 1.17\, \mu{\rm eV}~\sqrt{n_e/(10^9~{\rm cm}^{-3})},$  the detectable signal from ALP-photon conversion falls in the radio band and may be detected with radio interferometers.
The flux density at Earth is
\begin{equation}
S=\int\,\frac{d\Omega}{4\pi\,\Delta\nu}\,\rho\,v\,P_{a\rightarrow \gamma}\,e^{-\tau}\enspace,
\end{equation}
where $\rho$ is the DM density at the conversion location, $\tau$ is the photon optical depth, and $\Delta\nu$ is the bandwidth of the signal.

The optical depth at radio frequency, and for magnetic field strengths typical of the quiet corona, is dominated by free-free absorption, also called inverse-bremsstrahlung. This absorption channel affects larger ALP masses more significantly.
In addition, above sunspots, the cyclotron frequency $\Omega_B=eB/m_e$ may be of the same order as the ALP mass. In this case, gyro-resonance absorption occurs if $\omega=k\,\Omega_B,$ at some point along the photon trajectory. The integer $k$ refers to the different energy levels. We find gyro-resonance absorption to be significant up to $k=4$.

Taking these facts into account, and considering the capabilities of the upcoming radio interferometer SKA-1~\cite{SKA:2019}, we obtain the forecast shown in Figure~\ref{fig:res}.

%%%%%%%%%%%%%%%%%%%%%%%%%%%%%%%%%%%%
\section{Conversion of Solar Axions}\label{sec:nustar}
\begin{figure}[t]
    \centering
    \includegraphics[width=0.65\textwidth]{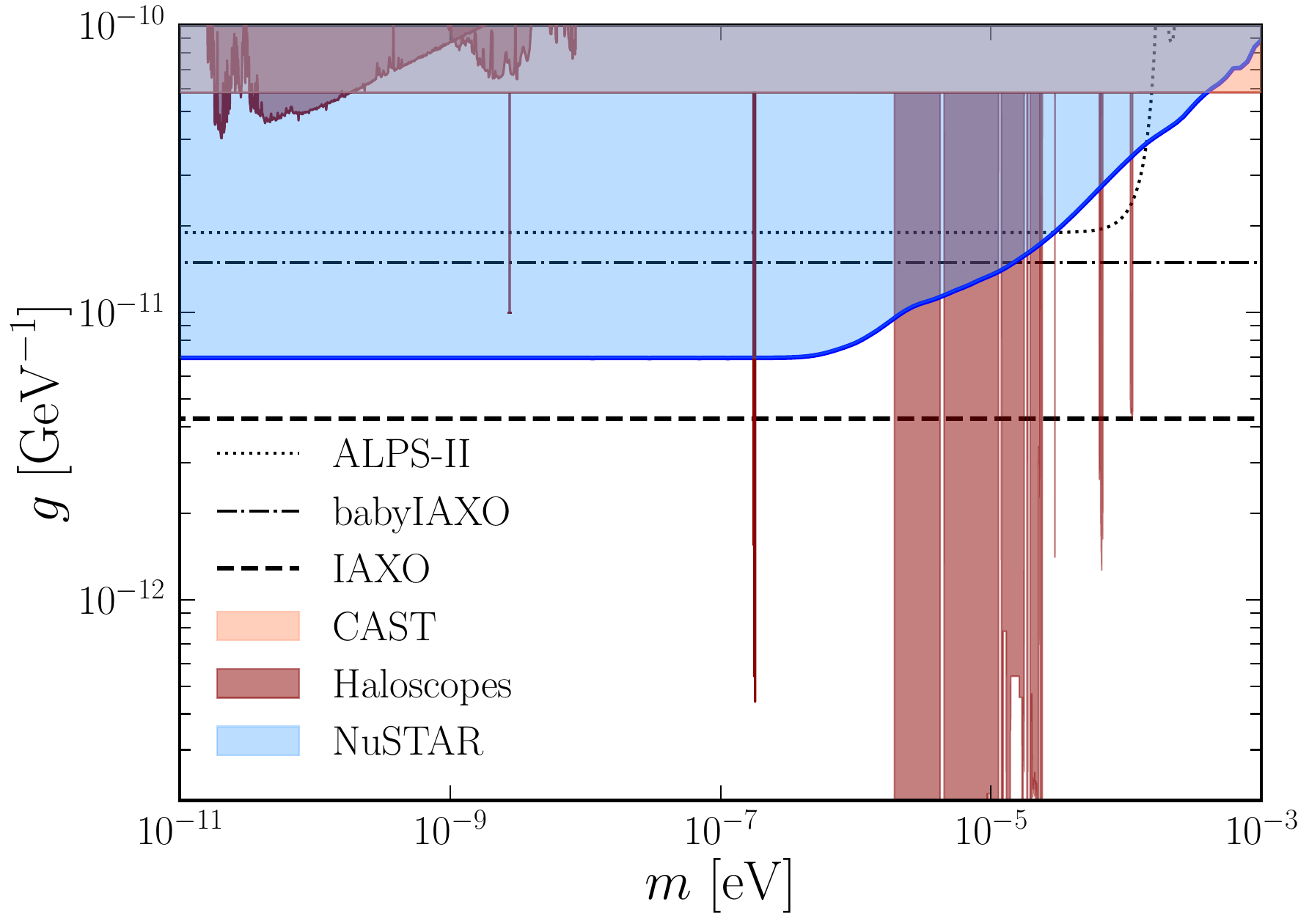}
    \caption{NuSTAR's 95\% CL upper limit on the axion–photon coupling strength $g$ from~\cite{Ruz:2024gkl} (blue region).
    Bounds from haloscopes~\cite{Asztalos_2010,Du_2018, Braine_2020, Bartram_2021, Boutan_2018, Bartram_2023, PhysRevD.42.1297, Zhong_2018, Backes_2021, haystaccollaboration2023new, Lee_2020, Jeong_2020, Kwon_2021, Lee_2022, Kim_2023, Yi_2023, Yang_2023, kim2023experimental, ahn2024extensive, Quiskamp_2022, quiskamp2023exclusion, McAllister_2017, Alesini_2019, Alesini_2021, Di_Vora_2023, Devlin:2021fpq,Ouellet_2019, Salemi_2021, Crisosto:2019fcj} are shown in dark red while the CAST helioscope constraint~\cite{CAST_PRL04, Andriamonje_2007, cast_jcap2009,CAST_PRL2011, PhysRevLett.112.091302, CAST_nature} is in light red. Sensitivity projections for future experiments~~\cite{RBahre_2013,PhysPotIAXO,Abeln:2021,Armengaud:2014gea} are also shown.
    Figure adapted from~\cite{Ruz:2024gkl}.
   }
    \label{fig:nustar}
\end{figure}

Solar ALPs produced in the core travel outward and interact with the magnetic field of the solar atmosphere, leading to an X-ray emission detectable by instruments like the NuSTAR telescope. During the solar minimum of 2020, NuSTAR observed the center of the solar disc for about 7 hours. Thanks to these data, new limits in $g$ are derived in~\cite{Ruz:2024gkl}, surpassing current and some future limits from laboratory experiments. The bound on $g$ is shown in Figure~\ref{fig:nustar}. 

Accurate modeling of the solar atmosphere is necessary to evaluate the conversion probability Eq.~\eqref{eq:prob}. This is particularly true for the magnetic field, which constitutes the major source of systematic uncertainty. We use state-of-the-art magnetohydrodynamics simulations for the chromospheric~\cite{Rempel_2014} and coronal~\cite{ps2017} magnetic field. We interpolate between the regions of validity of the two models. The perpendicular component of the magnetic field is shown in Figure 3 of~\cite{Ruz:2024gkl}. The model used for the coronal magnetic field, dominating the conversion probability for low axion masses, corresponds to the Sun's configuration during the total eclipse on July 2nd, 2019, while the NuSTAR observations were acquired on February 21st, 2020. Therefore, to further check the accuracy of our magnetic field model, we compare it with the Potential-Field Source-Surface (PFSS) model for the day of observation and find excellent agreement (see Figure 11 of~\cite{Ruz:2024gkl}). 

Taking into account uncertainties on the magnetic field, solar ALP flux, and X-ray background, we estimate the total systematic uncertainty on the bound to be less than 30\%. Such a level of uncertainty, mainly due to the magnetic field, is extremely low for an astrophysical ALP search and is possible thanks to our detailed knowledge of the solar atmosphere. Other astrophysical limits in the same region of parameter space~\cite{Li_2021, Li_2022, Reynolds_2020, Abramowski_2013, Ajello_2016, Davies_2023, Jacobsen:2022swa,Li:2024zst,Xiao:2020pra,Ning:2024eky,Dessert_2022_1, Dessert_2022_2,Noordhuis:2022ljw,Dessert_2020,Hoof_2023,Manzari:2024jns, Foster:2022fxn, Battye:2023oac,Escudero:2023vgv} typically come with larger uncertainties, due to the difficulty of modeling  ALP production and conversion in far-away astrophysical environments.

%%%%%%%%%%%%%%%%%%%%%%%%%%%%%%%%%%%%
\section{Conclusions}
The Sun provides a natural laboratory for studying axion-like particles, thanks to its magnetic fields and plasma. Its proximity implies that our knowledge of the solar environment is more detailed and solid than that of far-away astrophysical objects, allowing for robust bounds to be established.

In this report, we summarized two works that analyzed the signal from ALP-photon conversion for dark matter ALPs~~\cite{Todarello:2023ptf} and solar ALPs~\cite{Ruz:2024gkl}. 
These analyses demonstrate that ALP-photon conversion in the solar atmosphere can produce detectable signals in the radio and X-ray bands, respectively. Thanks to data from the NuSTAR X-ray telescope, the bounds shown in Fig.~\ref{fig:nustar} were derived.

Future advancements in solar modeling, thanks for example to the incoming 
new data from the Parker Solar Probe, and future X-ray observations, ideally concomitant with a solar eclipse, will help reduce the systematic uncertainty, and potentially strengthen, the bound of~\cite{Ruz:2024gkl}.

\section*{Acknowledgements}

The author thanks her collaborators who contributed to the work presented here: Marco Regis, Marco Taoso, Maurizio Giannotti, Jaime Ruz, Julia Vogel,  Brian Grefenstette, Hugh Hudson, Iain Hannah, Igor Irastorza, Chul Soo Kim, Thomas O'Shea, David Smith, and Javier Trujillo Bueno. ET is supported by the STFC Consolidated Grant [ST/T000732/1]. This article is based on the work from COST Action COSMIC WISPers CA21106, supported by COST (European Cooperation in Science and Technology).

%%%%%%%%%%%%%%%%%%%%%%%%%%%%%%%%%%%%
\bibliographystyle{JHEP}
\bibliography{refs}

\end{document}